\documentclass[prd,onecolumn,floatfix,notitlepage,12pt,tightenlines]{revtex4-1} 

\usepackage{amsmath}
\usepackage{graphicx}
\usepackage{dcolumn}
\usepackage{bm}
\usepackage{epsfig}
\usepackage{amssymb,latexsym,mathrsfs}
\usepackage{graphicx}
\usepackage{color}
\usepackage{hyperref}

\usepackage[utf8]{inputenc}
\usepackage{enumitem,amssymb}
\usepackage{ragged2e}
\newlist{thematic}{itemize}{8}
\setlist[thematic]{label=$\square$}
\usepackage{pifont}
\newcommand{\cmark}{\ding{51}}%
\newcommand{\done}{\rlap{$\square$}{\raisebox{2pt}{\large\hspace{1pt}\cmark}}%
\hspace{-2.5pt}}

\hypersetup{
    colorlinks=true,
    linkcolor=red,
    citecolor=blue,
} 

\newcommand{\be}{\begin{equation}}
\newcommand{\ee}{\end{equation}}
\newcommand{\bs}{\begin{split}} 
\newcommand{\bea}{\begin{eqnarray}}
\newcommand{\eea}{\end{eqnarray}}

\newcommand{\wn}{cm$^{-1}$}	
\newcommand{\about}{$\sim$}	

\begin{document}


\thispagestyle{empty} 

\huge
\begin{center} 
Astro2020 Science White Paper
\end{center} 
\vspace{1cm} 
\begin{flushleft} 
Direct Acceleration: Cosmic and Exoplanet Synergies\\  
\end{flushleft} 

\vspace{2cm} 

\normalsize

  
 \noindent \textbf{Thematic Areas:} \hspace*{10pt} $\done$ Planetary Systems \hspace*{10pt} $\square$ Star and Planet Formation \\ 
$\square$ Formation and Evolution of Compact Objects \hspace*{10pt} $\done$ Cosmology and Fundamental Physics\\ 
$\square$  Stars and Stellar Evolution \hspace*{1pt} $\square$ Resolved Stellar Populations and their Environments\\ 
$\square$    Galaxy Evolution   \hspace*{10pt} $\square$             Multi-Messenger Astronomy and Astrophysics\\ 
\\ 

\noindent\textbf{Principal Author:}

\noindent Name:   David Erskine\\                                             
Institution:  Lawrence Livermore National Laboratory\\ 
Email: erskine1@llnl.gov\\ 
 
\noindent\textbf{Co-authors:}\\ 

\noindent Malte Buschmann, University of Michigan\\ 
Richard Easther, University of Auckland\\ 
Simone Ferraro, Lawrence Berkeley National Laboratory\\ 
Alex Kim, Lawrence Berkeley National Laboratory\\
Eric Linder, University of California Berkeley\\ 
Philip Muirhead, Boston University\\ 
David Phillips, Harvard Center for Astrophysics\\ 
Aakash Ravi, Harvard Center for Astrophysics, Harvard University\\ 
Benjamin Safdi, University of Michigan\\ 
Emmanuel Schaan, Lawrence Berkeley National Laboratory\\ 
Hamish Silverwood, University of Barcelona\\ 
Ronald Walsworth, Harvard Center for Astrophysics 

\newpage

\setcounter{page}{1}

\title{Direct Acceleration: Cosmic and Exoplanet Synergies}

\begin{abstract} 
Direct measurement of acceleration is a key scientific goal for both 
cosmology and exoplanets. For cosmology, the concept of redshift drift 
(more than 60 years old by the 2020s) could directly establish the 
Friedmann-Lema{\^\i}tre-Robertson-Walker model. It 
would increase the dark energy figure of merit by a factor of 3 beyond 
Stage 4 experiments, in combination with cosmic microwave background 
measurements. For exoplanets, the same technology required provides 
unprecedented radial velocity accuracy, enabling detection of Earth mass 
planets in the habitable zone. Other science cases include mapping the Milky Way gravitational potential and testing its dark matter distribution.  
\end{abstract} 


\maketitle

\section{Introduction} 

Four frontier science areas can make great strides with the development of 
highly accurate and stable spectroscopy: 
\begin{itemize} 
\setlength\itemsep{0.1cm} 
\item Cosmic redshift drift and direct detection of cosmic acceleration;  
\item Earth mass exoplanet detection from radial velocities;  
\item Milky Way structure mapping through stellar accelerations;  
\item Dark matter properties through Milky Way gravity mapping.   
\end{itemize} 
Redshifts and Doppler velocities are central measurements for cosmology 
and for exoplanets respectively. While greater accuracy in exoplanet 
radial velocities is a recognized goal and major technology driver, in 
order to pick out Earth mass planets around distant stars, cosmological 
redshift precision has been sufficient for standard needs. However, the 
concept of redshift evolution over long timescales due to cosmic expansion 
-- redshift drift -- has repeatedly been brought up in the literature since 
it was first introduced by McVittie and Sandage in back to back articles in 
1962 \cite{citeA,citeA2}. 

Just as redshift is a direct measure of cosmic expansion, so redshift drift 
is a direct measure of cosmic acceleration. The characterization and 
identification of the origin of current cosmic acceleration is a major 
goal of cosmic surveys in the 2020s, such as the Large Synoptic Survey 
Telescope (LSST) and Dark Energy Spectroscopic Instrument (DESI). This 
cosmic acceleration is due to unknown new energy in the universe -- a 
cosmological constant or dynamical dark energy -- or to new laws of 
gravity beyond Einstein's general relativity -- or to a breakdown in the 
homogeneous and isotropic spacetime governed by the equation of the 
Friedmann-Lema{\^\i}tre-Robertson-Walker (FLRW) model. Direct measurement 
of cosmic acceleration can rule out the last possibility by confirming the 
reality of dynamical acceleration and give incisive new constraints on the 
properties of the effective dark energy, greatly complementary to those 
coming from the 2020s experiments under construction. Indeed it could 
increase their probative power by a factor three, turning them into Stage 5 
experiments. 

Redshift is due, in part, to the Doppler effect of the
relative velocity 
between the source and observer.
The change in ``peculiar'' velocity of a source induced by local
forces thus represents
a different contribution to redshift drift.  While this motion might be considered as noise
when determining the overall evolution of cosmic expansion, upon closer inspection it is an
interesting signal of the local gravitational forces caused by (dark) matter
and is thus a probe of the mass distribution within the Milky Way,
globular clusters, and galaxy clusters. Acceleration measurements of 
stars within the Milky Way map out the Galactic potential and explore the 
distribution and particle nature of dark matter \cite{citeB,citeC}. 

The technology path for high accuracy exoplanet radial velocity 
measurements is also eminently suited for cosmic acceleration observations 
\cite{RedshiftWhirl2014}. Recognition of better astrophysical line sources 
-- emission line galaxy surveys, particularly of OII doublets allowing 
differential rather than absolute measurements -- a better window for dark 
energy characterization -- at low redshift not high redshift -- as well as 
new methods for canceling instrumental systematics, argues that the 2020s 
is the decade to use this synergy between cosmology and exoplanets to push 
development toward both science goals.

\section{The Accuracy Challenge} 

An Earth mass exoplanet orbiting a solar mass star in the habitable 
zone for life induces a radial velocity signal on the star of a few to 
ten cm/s. This is a Doppler shift of order $v/c\approx 10^{-10}$. 
The cosmological redshift of a source (say a galaxy emission line) changes 
by order one in a Hubble time, or approximately a factor $\Delta z\approx 
10^{-10}$ in one year. 
These are extremely challenging levels, but their matching size reveals the 
synergy that exists in these science goals from common technology 
development. 

In the cosmological case, we could build up measurements over 
a longer baseline, say of 10 years, or alternatively produce not just a 
detection, but a 
higher signal to noise characterization. So a  1\% measurement 
over 10 years requires $10^{-11}$ accuracy.  (We prefer $10^{-12}$ accuracy.  Since one can improve by $\sqrt{N}$ 
 the precision using $N$ sources, we will phrase this as $10^{-11}$ 
accuracy.) Large mass stars 
having Earth-like exoplanets further out have smaller Doppler shifts, so $10^{-11}$ 
accuracy benefits exoplanetary discovery and characterization science as well. 
Current Doppler radial velocity measurement precision has broken through the \about 1~m/s level and is poised at  
$10^{-9}$ \cite{Fischer2016PASP}. Concerted efforts for future improvement, and 
dedicated telescope time, could pay large 
scientific dividends for both cosmology and exoplanets search and characterization.

\section{Scientific Outcomes} 

Let us examine the prospective outcomes of acceleration measurements at 
the desired level. 
First for cosmology is simply the detection of the redshift drift. At 
high redshifts the Hubble time is shorter so the redshift drift is larger, 
though negative due to the matter dominated deceleration. This has led 
many to propose using sources at redshift $z>2$. However, this does not 
account for the science goals -- verification of the FLRW model within the 
accelerating epoch and characterization of the dark energy properties. 
The redshift drift $\dot z=dz/dt_0=H_0\,(1+z)-H(z)$ peaks in acceleration 
around $z\approx0.9$, achieving half its maximum height already by $z=0.25$. 
Furthermore, at low redshift is the greatest discriminating power for 
dark energy.

\begin{figure}[ht!] 
\begin{minipage}[c]{0.5\textwidth} 
\includegraphics[width=\textwidth]{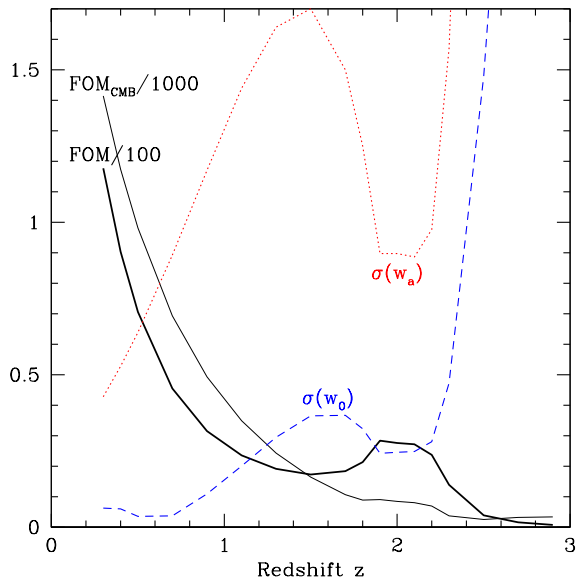} 
\end{minipage}\hfill
  \begin{minipage}[c]{0.5\textwidth} 
\caption{
Constraints at $1\sigma$ on dark energy $w_0$ and $w_a$, and their joint 
figure of merit (FOM), are plotted vs central redshift for experiments 
consisting of five measurements of redshift drift at 1\% precision. CMB 
constraints are included in (only) the FOM$_{\rm CMB}$ curve; note it is 
shown divided by 1000 (i.e.\ the maximum is 1400), rather than 100 like 
the FOM curve without CMB. 
} 
\label{fig:fom} 
\end{minipage} 
\end{figure}

Figure~\ref{fig:fom} shows the Dark Energy Task Force figure of merit on 
the dark energy equation of state parameters that distinguish between 
classes of cosmic acceleration: the present value $w_0$ and 
a measure of its time variation $w_a$. 
Low redshift measurements around $z=0.3$, in combination with CMB data 
already existing, deliver a figure of merit of 1400. Note this is independent 
of all other Stage 4 dark energy experiments such as LSST and DESI, offering 
not only a factor of 3 gain over them in isolation, and an independent crosscheck, but 
further gains from combination of all data together. Furthermore note that 
the emission line galaxies to be targeted are useful in themselves to the 
DESI and LSST surveys. Even diluting to a 5\% redshift drift precision 
provides a roughly equal crosscheck to a Stage 4 experiment. Moreover, 
an improved measurement of the Hubble constant to 1.4\% precision can increase 
the FOM to 2300, showing further synergy in 2020s science goals. 

By contrast, surveys aiming at $z>2$ are near pessimal. They are aiming at 
simple detection, but have little leverage on dark energy at such high 
redshift. This is the strategy followed by the CODEX spectrograph proposed 
for the European Extremely Large Telescope (EELT). By using many 
Lyman-$\alpha$ lines in quasar absorption spectra they hope to reduce the 
requirements on measurement precision. However, this drives them to very 
high resolution ($R>120,000$) and a poor redshift range, as well as laying 
them open to astrophysical systematics from gas velocities and varying 
ionizing radiation field. 

Thus the optimal low redshift range is ripe for US endeavors. Again, this 
is optimal not only for dark energy properties but in testing the FLRW 
framework against, e.g.\ void models or inhomogeneous universes such as 
Lema{\^\i}tre-Tolman-Bondi or Szekeres models that give the mirage of 
acceleration without true dynamics. 

A further breakthrough involves the use of emission line surveys focusing 
on the forbidden OII doublet. While redshift (and redshift drift) affects 
the frequency of a line, it equally affects the spacing between lines. Thus 
we can turn an absolute measurement into a differential  
measurement of the 
spacing between well known doublet lines whose properties are determined by 
atomic physics. Emission line surveys using OII are standard workhorses of 
cosmology, used in the BOSS, eBOSS, and DESI surveys, and need only 
spectrographs of modest resolution $R\approx5000$. 

Galaxies have spatial structure with
internal dynamics.  Integral Field Unit (IFU) spectroscopy
can provide spatial resolution, not only to distinguish between
bulk and internal velocity evolution, but also to take advantage
of the multiple spatially-resolved measurements of a line, each
of which is narrower than the line when spatially-unresolved.

\section{Systematics Mitigation} 

The chief instrumental challenge is not spectral resolution, but wavelength stability: the point spread function (PSF) of conventional spectrographs drifts in position and shape under duress from thermal changes to the diffraction grating, fluctuations of air internal and external to spectrograph, a changing pupil,  flexure of optical fibers etc.  Conventional means for mitigating PSF drift include thermal control, vacuum tanks, adaptive optics, fiber optic scramblers, and laser frequency comb calibrants.  These measures reduce the ``insult'', $\delta \lambda_{\rm insult}$, which reduces the error in the final spectrum. For purely dispersive spectrographs,
$\delta \lambda_{\rm final} = \delta \lambda_{\rm insult}$, since the spatial scale of the detector is directly linked to the final spectrum, so requirements are 
severe on $\delta \lambda_{\rm insult}$.

While such mitigations should be used if affordable, dispersive spectrograph stability can be further improved by 2 or 3 orders of magnitude by inclusion of a Michelson interferometer in series, as in externally dispersed interferometry (EDI).  This technique has been used for precision Doppler radial velocimetry and high resolution spectroscopy \cite{ErskineGe2000,ErskineEDItheory2003,ResBoostApJ2003,MuirheadPASP2011,TediTenx2016part1,TediTenx2016part2}  
and discovered an exoplanet around star HD 102195 \cite{VirgoGe2006}.
In this method the detailed wavelength determination is decoupled from the spatial scale in the disperser and its drift $\delta \lambda_{\rm insult}$.  Instead, the detailed wavelength is determined by the phase of a fringe (intensity measurement) in an interferometer cavity, which is calibrated by a spectral reference such as an iodine cell, ThAr lamp, or laser frequency comb.  The cavity PSF is sinusoidal and has only three degrees of freedom (amplitude, period, phase).  This is much easier to control or calibrate than the hundreds of degrees of freedom for a disperser PSF (at least one per grating groove).  

\begin{figure}[htbp!]
\includegraphics[width=0.47\columnwidth]{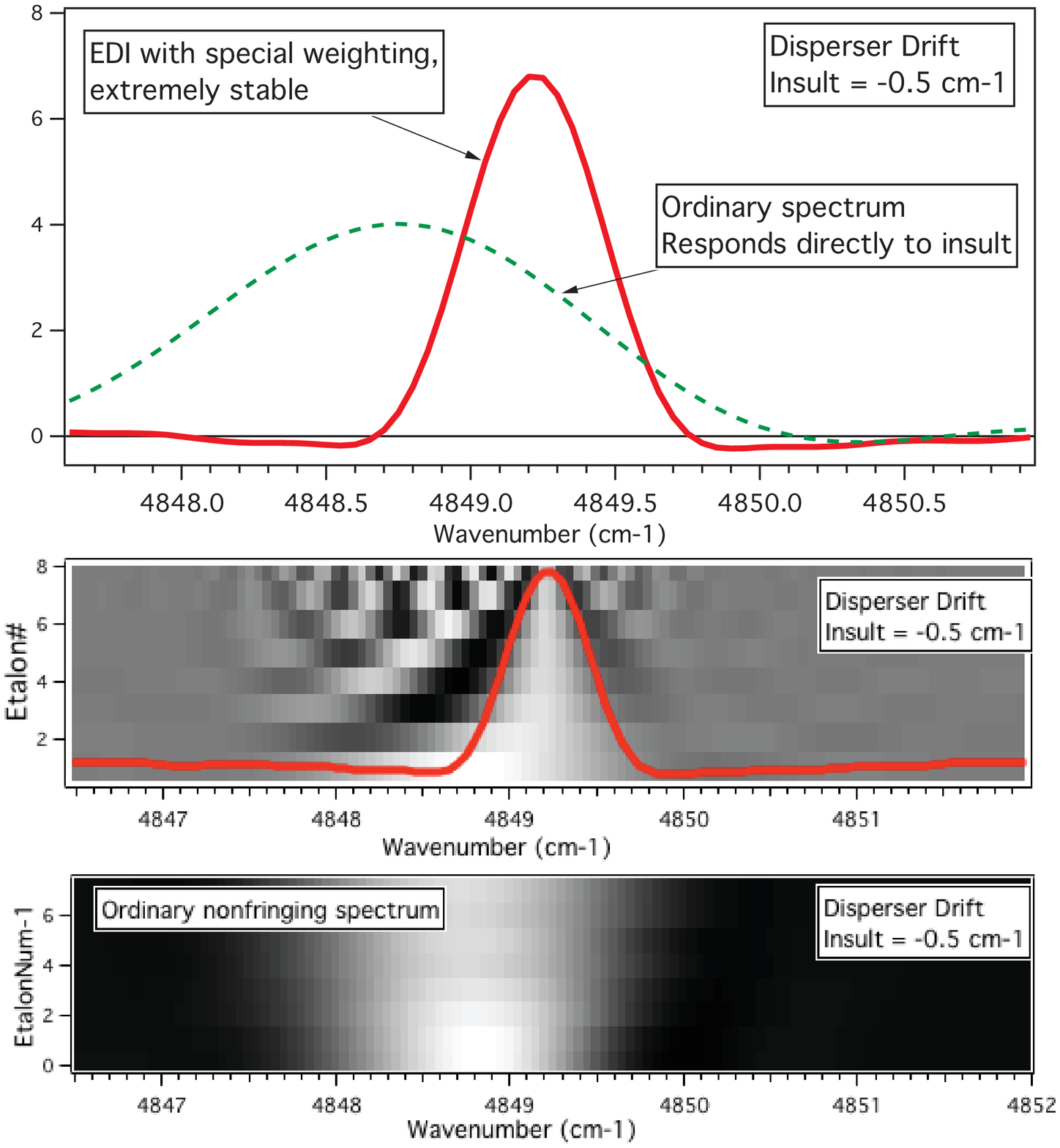}
\includegraphics[width=0.47\columnwidth]{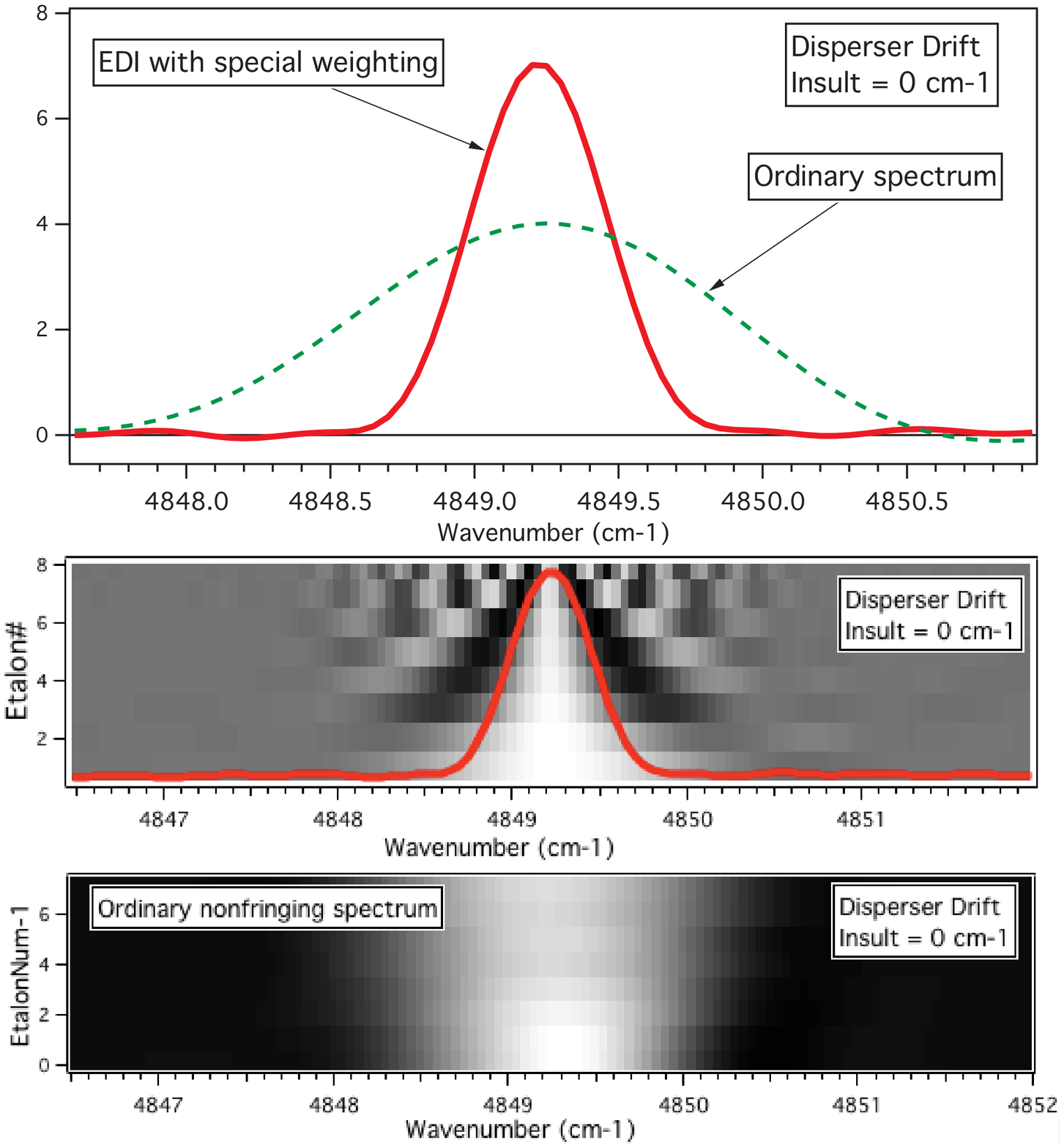}
\caption{
A multiple delay externally dispersed interferometer boosts the stability and resolution of the disperser in series with it.  Precision wavelength is obtained from the interferometric fringe phase, not the disperser, which mainly affects the fringe envelope (middle panels).  Multiple delays (etalons) are used having different periodicities, which are summed to form the net EDI peak (red curve).  Mount  Palomar Observatory ThAr lamp line data \cite{TediTenx2016part1} (right panels)  is artificially shifted on the detector by 0.5 \wn\ (left panels).  The disperser peak (green dashes) shifts directly, while the net EDI peak moves only 1/1000$^{\rm th}$ the amount.
} 
\label{fig:calib} 
\end{figure}

Hence now the final PSF drift is given by
$\delta \lambda_{final} = \delta \lambda_{insult} * TRC$, 
where TRC is the translation reaction coefficient of the spectroscopic method used.  For dispersive spectroscopy TRC = 1; but for EDI using multiple delays we show (Sec.~10 of Ref.~\cite{TediTenx2016part1}) how to theoretically make TRC = 0.  In recent demonstrations on a single ThAr line using the same data analysis software used in observations, we have obtained TRC as small as 1/1000! In our previous work we showed that the benefit of multiple delays, but without special weights, can achieve TRC \about 1/20 \cite{TediTenx2016part1}.  In recent work however we used a technique called ``crossfading" using strategically chosen weightings to force phase cancellation between a pair of delays, and have improved stability to TRC \about 1/1000 (Fig.~\ref{fig:calib}).

The crossfading technique works because under the same detector wavelength drift, EDI signals measured by a high delay will twist in one direction, and by a low delay will twist in the opposite direction.  For final spectrum frequencies that lie in between a pair of (overlapping) delays, we choose weightings that cancel the net phase shift.  We reweight (in software) for every pair of delays for each frequency, up to a limiting frequency of the highest delay.  

Crossfading also works with a single delay that overlaps the native PSF, 
and this  stabilizes against drifts of all time scales.  We have verified 
that crossfading also stabilizes against changes in PSF width or 
asymmetry. 

In summary, the science cases can gain by significant instrumental 
systematics mitigation using crossfading EDI, because spatial drift of the disperser PSF is a dominant error source for current radial velocity spectrographs. Stability benefits multiply, so  
if using a fiber scrambler or vacuum tank provides, say, a 10$^4$ PSF drift reduction, then including the EDI could produce a 10$^4\times$1000 = 10$^7$ net reduction.

\section{Conclusion} 

Direct acceleration measurements are science goals of both cosmology, in 
the form of redshift drift, and exoplanet research, in high precision radial 
velocities to find Earth mass exoplanets, as well as Milky Way and dark matter mapping science cases. Support for technology advances 
in application of interferometric spectroscopy, and survey time, for 
these science goals can lead to revolutionary results. For cosmology, the 
use of differential measurements of forbidden OII doublet lines at low 
redshift is the optimal path, one not yet taken. Exposure time calculations 
indicate precision of a few times $10^{-9}$ can be achieved in 8 hours on a 
10 meter telescope, for a single bright emission line galaxy at low redshift. 
Large numbers of objects will enable redshift drift measurements below the 
$10^{-10}$ level, with development of improved instrumental methods such as 
the already demonstrated delay shift technique. Astrophysical, ``peculiar 
accelerations'' will also be mapped by such surveys, revealing structure 
in the dark and dynamic universe, as well as our own Milky Way galaxy. 
The LIGO gravitational wave interferometer is a clear example of the advantages of using fringe phase shifts for making precision measurements for astrophysics. 

The synergy of the enabling technology of high accuracy and stability 
spectroscopy for four astrophysics frontiers is truly extraordinary, 
and the 2020s is the decade when, with clear support, it can reach 
maturity. 

\clearpage 


\end{document}